\documentclass[a4paper, oneside, twocolumn, notitlepage, 10pt]{extarticle_ecoc}
\usepackage{ecoc}
\usepackage{color,soul}
\begin{document}
\selectlanguage{english}    


\title{Power Evolution Prediction and Optimization in a Multi-span System Based on Component-wise System Modeling}%


\author{
    Metodi P. Yankov\textsuperscript{(1)}, Uiara Celine de Moura\textsuperscript{(1)}, Francesco Da Ros\textsuperscript{(1)}
}

\maketitle                  


\begin{strip}
 \begin{author_descr}

   \textsuperscript{(1)} Department of Photonics Engineering, Technical University of Denmark, 2800 Kgs. Lyngby, Denmark,
   \textcolor{blue}{\uline{meya@fotonik.dtu.dk}}

 \end{author_descr}
\end{strip}

\setstretch{1.1}


\begin{strip}
  \begin{ecoc_abstract}
    Cascades of a machine learning-based EDFA gain model trained on a single physical device and a fully differentiable stimulated Raman scattering fiber model are used to predict and optimize the power profile at the output of an experimental multi-span fully-loaded C-band optical communication system.
  \end{ecoc_abstract}
\end{strip}


\section{Introduction}
Power evolution, alongside optical signal to noise ratio (OSNR) and electrical SNR monitoring and optimization in wavelength division multiplexed (WDM) optical communication systems is an integral part of network operations as it is one of the most critical performance indicators. Predicting the power profile at the output of a link allows for software defined networks to automatically allocate resources, specify physical layer transmission parameters (such as modulation format and forward error correction rate) in each channel and thus increase the rate and energy efficiency of the network. Optimization of said profile would allow to increase the efficiency even further by for example ensuring uniform quality of service to all users sharing a given fiber connection. The SNR profile can be predicted and optimized using a stimulated Raman scattering (SRS)-appended Gaussian noise model \cite{Roberts}. The model \cite{Roberts} is fully theoretic and assumes ideal erbium doped fiber amplifiers (EDFAs) with flat gain and noise figure profiles. This assumption is often too strong even for EDFAs with gain-flattening filters (GFFs) \cite{EDFA_gain_JLT}. Furthermore, while EDFAs with GFFs enable long-haul transmission with relatively stable power evolution, they are by definition energy wasteful. Machine learning (ML) is used to model the EDFA gain tilt \cite{Maria, EDFA_gain_JLT, Zhu:20} and predict and optimize the power and SNR evolution through a simulation-based point-to-point link \cite{Maria}. The model is trained on a single physical device and assumed to generalize to the rest of the devices in the link, which is not demonstrated. Furthermore, linear region of operation is assumed, where the SRS effect is not pronounced. The generalization problem can be treated instead by modeling the entire link, and then using the model to predict and optimize the OSNR \cite{JuhnoCho}. This approach is also limited to the linear region of transmission. Furthermore, a separate ML model is needed for each desired link configuration, which requires many hours of training data generation and cannot be performed in real time. 

In this paper, cascades of an EDFA gain ML-based model and a fiber model are used to model a multi-span system with up to three different physical spans. The EDFA model is trained on a single physical device. The fiber model takes into account the SRS. Modeling error is achieved of less than 0.6 dB\textsuperscript{2} of mean squared error (MSE), and less than 2 dB excursion when the system input power is optimized for flatness at the system output. To our knowledge, this is the first experimental demonstration of a component-wise model used for the prediction and optimization of multi-span system performance. 

\section{Theoretical model}
\begin{figure}[t]
   \centering
        \includegraphics[width=1.05\linewidth]{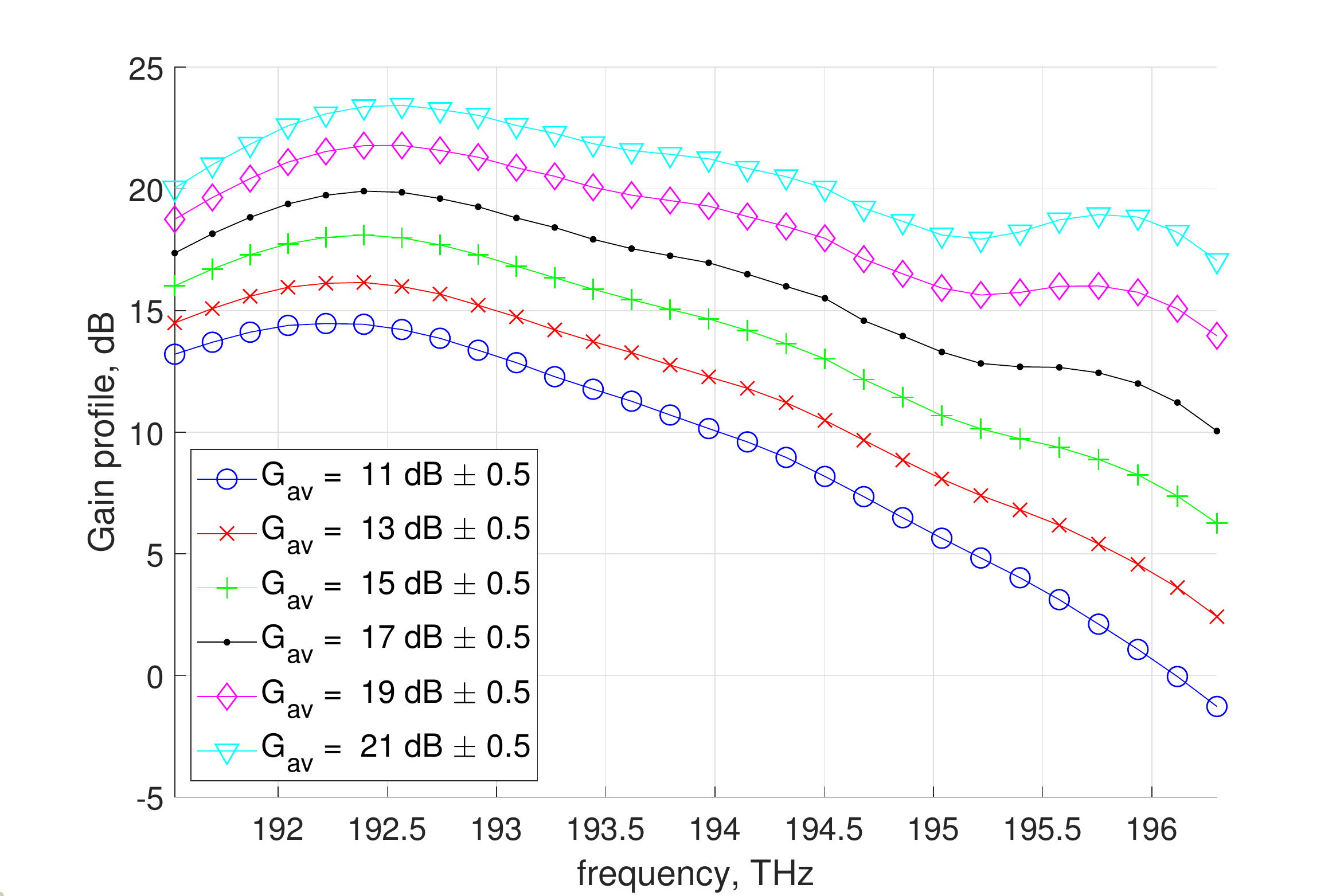}
    \caption{Gain profile of the EDFA used to obtain the ML model for different average gains. This example is at output power of 18 dBm.}
    \label{fig:EDFA}
\end{figure}

A general purpose neural network (NN) (two hidden layers, 256 and 128 nodes, respectively, with ReLU activation) is used to model the gain tilt of the EDFA as a function of the 1) normalized input power profile; 2) total input power; and 3) total output power\cite{DaRosECOC2020}. The C-band is discretized to 83 distinct uniformly spaced frequencies, making for a model input dimensionality of 85. The average gain profile of the EDFA is given in Fig.~\ref{fig:EDFA} for different average total gains and total output power of $18$ dBm. As demonstrated, the gain exhibits a strong excursion, more pronounced at low average gains which can be expected for this model of booster EDFA (Keopsys, KPS-STD-BT-C-18-SD-111-FA-FA). The model achieves an MSE$<$0.06 dB\textsuperscript{2} of cross-device gain prediction. Further details are omitted here and can be found in \cite{DaRosECOC2020}. Instead, the focus here is on system optimization through the multi-span  application of the model.

Two models are considered for the power evolution in the fiber: 1) a bulk, frequency-flat loss with loss coefficient of $\alpha=0.2$ dB/km; and 2) a model which includes the Raman gain and transfer of power from high frequencies to low frequencies. The latter is modeled by discretizing the fiber length into steps and calculating the Raman gain evolution per step as\cite{Roberts}
\begin{align}
\label{eq:SRS}
P_n(z) & = -\alpha L_s +  \\
       & \sum_{m=1}^{83} \frac{g_R(\omega_m-\omega_n)}{A_{eff}}L_{eff}(L_s)e^{P_m(z-L_s)}, \nonumber
\end{align}
where $P_n(z)$ is the power in the log-domain of the $n-$th carrier at distance $z$, $L_s=100$ m is the step size, $g_R$ is the tabulated Raman gain coefficient for a given frequency offset, $\omega_n$ is the angular frequency of the $n-$th carrier, $A_{eff}$ is the fiber effective area, and $L_{eff}(L) = \frac{1-\exp(-2\cdot\alpha\cdot L)}{ 2\cdot \alpha}$ is the effective power interaction length\cite{Roberts}. Standard single mode fiber parameters are assumed. 

\begin{figure}[t]
   \centering
        \includegraphics[width=1.0\linewidth]{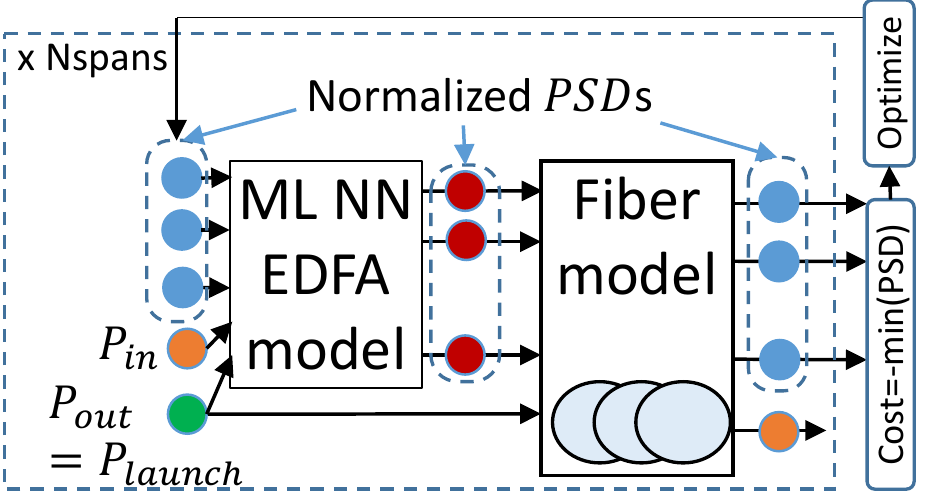}
    \caption{Optimization setup for a cascade of EDFA ML-based model and a theoretical fiber model. The total output power of the fiber model and the corresponding normalized power profile are the input to the EDFA model on the next span.}
    \label{fig:optimization}
\end{figure}

A cascade of SRS fiber models and ML EDFA gain models are then used to predict the power of a multi-span system, and then attempt to flatten the output power spectral density (PSD) $PSD_{out}$ using gradient descent. An illustration of the process is given in Fig.~\ref{fig:optimization}. The EDFAs are assumed to be controlled in constant output power mode, where the output power is set to the desired launch power $P_{launch}$. The input of every element in the cascade is a normalized PSD (such that maximum power per channel is 0 dBm) and total input power. The total input power of the first EDFA in the cascade is set to be $P_{launch} - L_{F1}\cdot \alpha$, where $L_{F1}$ is the length of the first fiber. The total input power of the subsequent EDFAs are governed by the launch power and the loss of the corresponding preceding fiber. The cost function used for the optimization is $Cost=-\min(PSD_{out})$, which for a normalized PSD is minimized with a flat profile. Automatic differentiation in PyTorch is then used to perform the optimization. 

\section{Experimental setup}
\begin{figure}[t]
   \centering
        \includegraphics[width=1.0\linewidth]{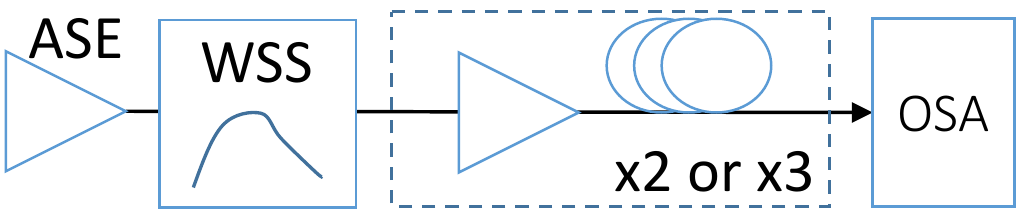}
    \caption{Experimental multi-span transmission system.}
    \label{fig:exp_setup}
\end{figure}

The multi-span experimental setup is given in Fig.~\ref{fig:exp_setup}. Amplified spontaneous emission (ASE) noise source is shaped using a wavelength selective switch (WSS). The WSS is first calibrated using pre-attenuation to flatten the ASE spectrum. The desired input power profile is then transformed into an attenuation profile and applied to the WSS in addition to the calibration profile, while also rescaling it to the desired total input power. Three different EDFAs of the same make and fiber spools of lengths 70, 80, 90 and 100 km are used in random sequence combinations to create two- and three-span inline amplified systems. In the 2-span cases, the EDFA used to train the ML model is \textbf{\textit{not}} employed. An optical spectrum analyzer (OSA) is used to capture the power profile at the output of the system. 

\section{Results}

\begin{figure*}[t]
   \centering
        \includegraphics[width=.32\linewidth]{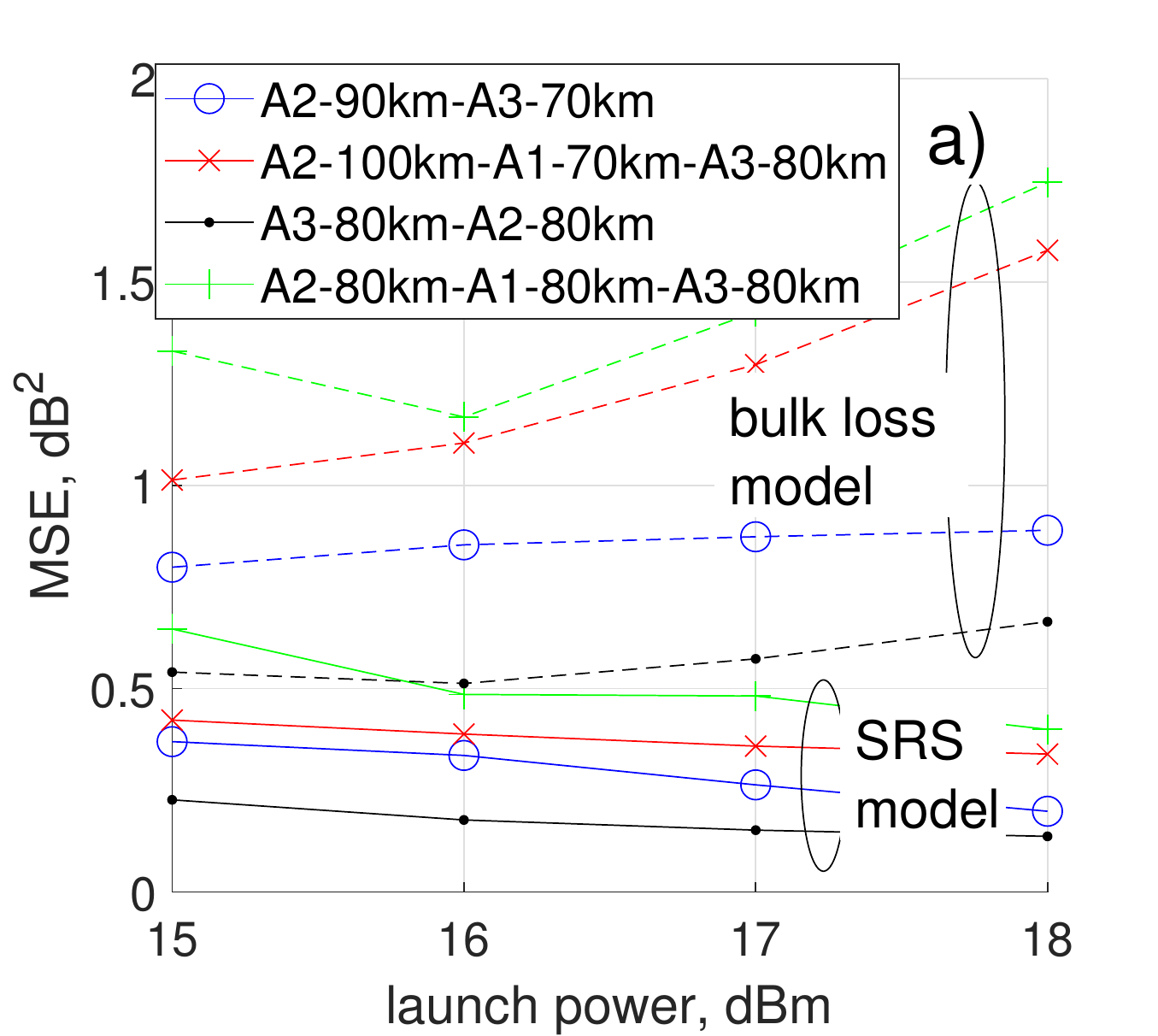}
        \includegraphics[width=.32\linewidth]{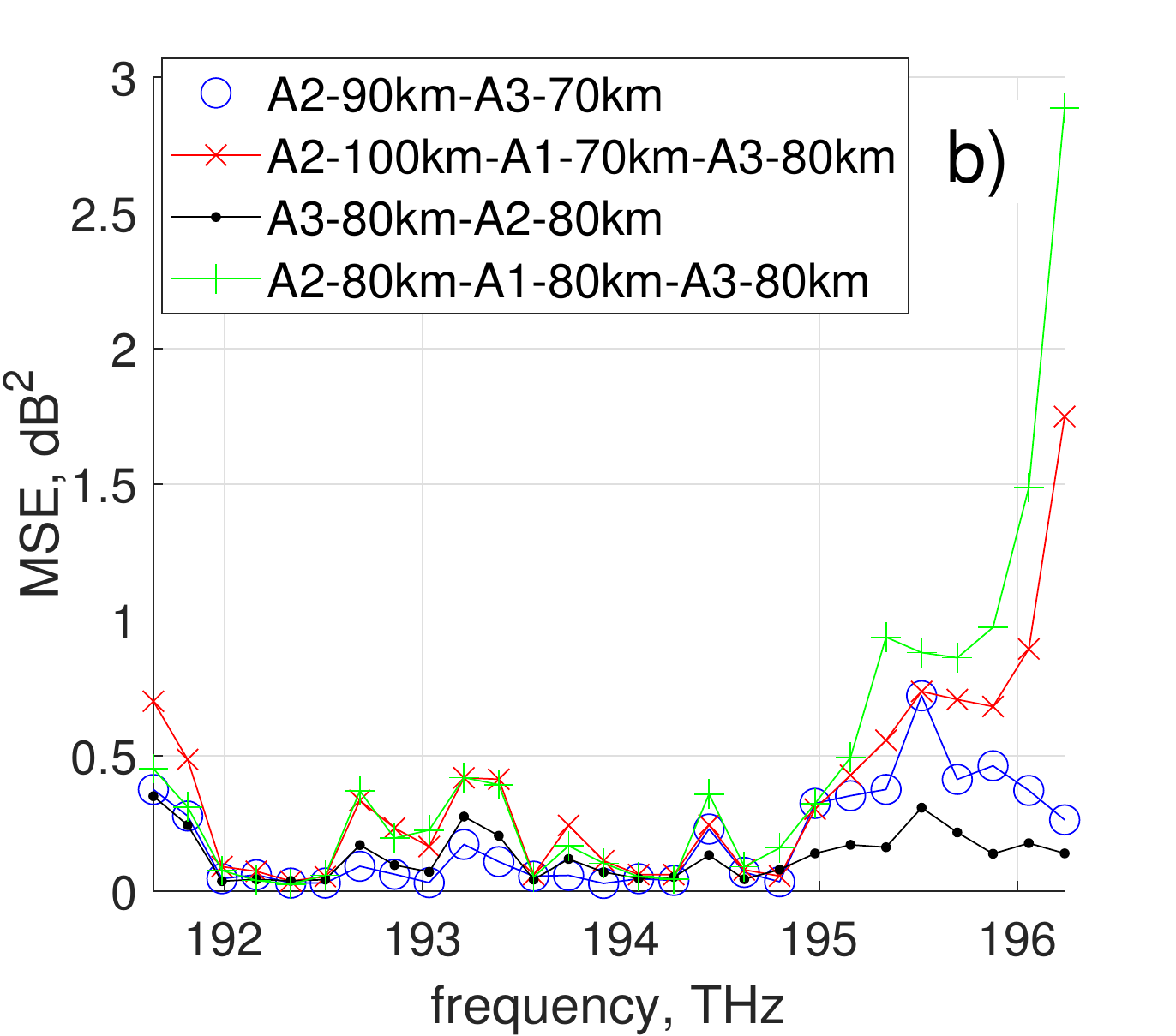}
        \includegraphics[width=.32\linewidth]{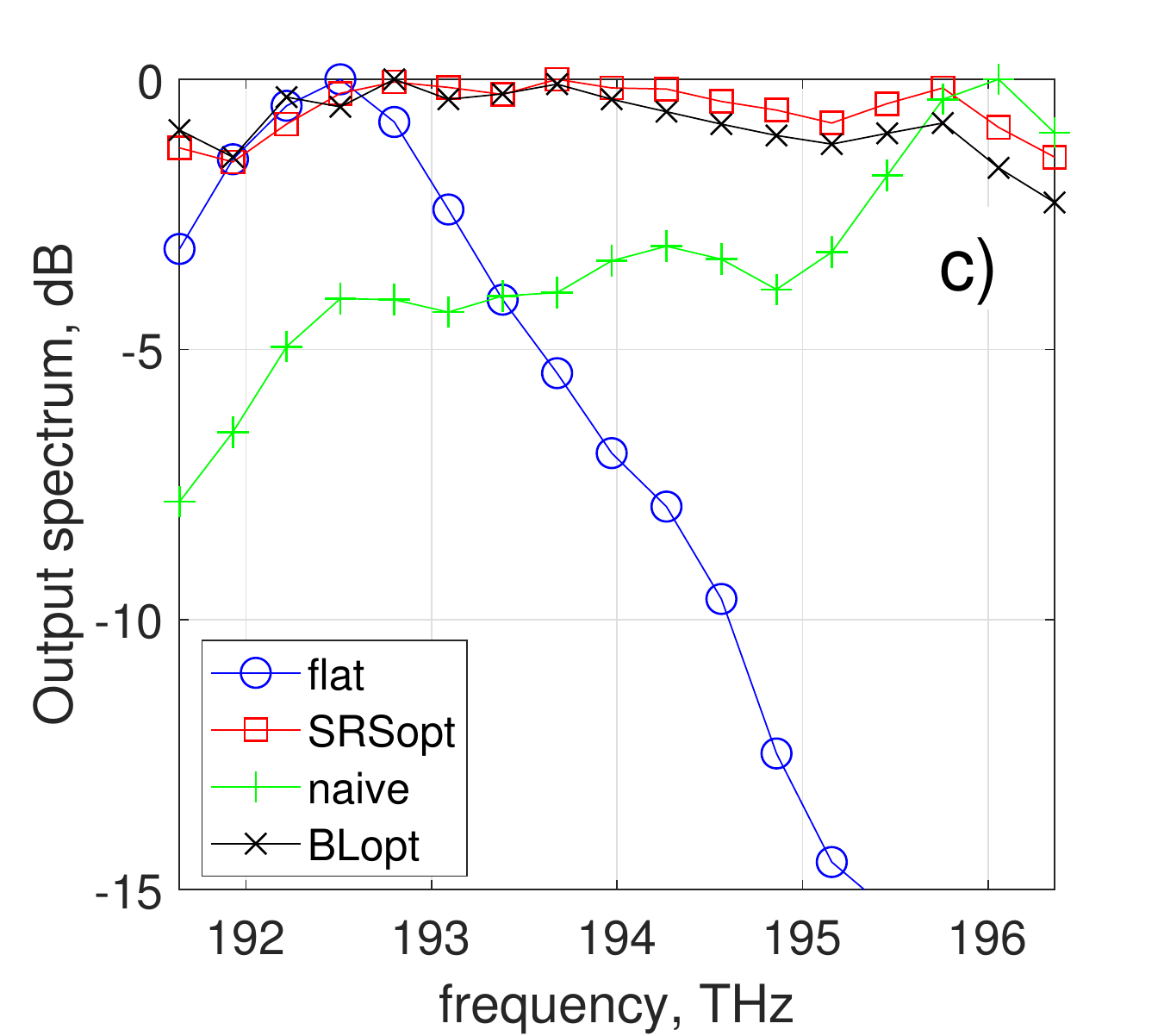}
    \caption{Modeling accuracy and optimization results. \textbf{a:} MSE at different powers for the two fiber models and four selected system configurations as given in the legends; \textbf{b:} MSE for different wavelengths at $P_{launch}=18$ dBm and SRS-included fiber model; \textbf{c:} Example \textbf{'A2-90km-A3-70km'} system \textit{response} to optimized and flat input profiles at $P_{launch}=18$ dBm.}
    \label{fig:results}
\end{figure*}

Testing data profiles are generated using Gaussian random walk $a_n = a_{n-1}+w_n$, where $a_n$ is the power at the $n-$th frequency carrier, $w_n$ is a zero-mean Gaussian variable with variance $\sigma^2_W$, and $a_0 \in [-14; 0]$. These profiles are then filtered by an all-ones filter of different lengths in order to obtain smooth profiles of higher and lower smoothness. A total of 2000 profiles with excursions ranging from 0 to 15 dB (controlled by $\sigma^2_W$) are generated and transmitted through the system. The performance prediction in terms of average MSE between the predicted and measured profiles is given in Fig.~\ref{fig:results}a). The system configuration order is specified in the legend, where A1, A2 and A3 are the employed amplifiers, A1 being the EDFA used for the ML model. The SRS-included model provides accuracy of less than 0.6 dB\textsuperscript{2} for the 3-span systems and better for fewer spans. The bulk loss model suffers high inaccuracy, especially at high launch powers where the SRS is more pronounced. The average accuracy is heavily impacted by the generally low accuracy at high frequencies (see Fig.~\ref{fig:results}b)), in turn attributed to the strong EDFA gain tilt and thus low received power at those frequencies, approaching the noise floor of the OSA. This is more pronounced for 3 spans due to the cumulative effect of the gain tilt. Furthermore, the EDFA model accuracy is lower at high frequencies\cite{DaRosECOC2020}. 

The block diagram from Fig.~\ref{fig:optimization} is then used to numerically optimize the input power profile so that flat output is achieved. The input profiles under test are: 1) optimized using the SRS model (SRSopt); 2) optimized using the bulk loss model (BLopt; 3) flat profile; and 4) a naively optimized profile. The last is obtained by inverting the system response to a flat input profile. Example responses of the experimental system to those input profiles are given in Fig.~\ref{fig:results}c) for the 2-span system of 90 km and 70 km per span, respectively. Due to the strong EDFA gain tilt, a flat input results in more than 15 dB of excursion already at the output of a simple 2-span system. Due to the EDFA gain and fiber nonlinearities, the naive optimization over-compensates for that, and results in an $\approx$8 dB tilt in the opposite direction. The profile optimized using the bulk loss model results in lower power at high frequencies and $\approx$3 dB excursion, which is then improved to 2 dB when optimization is performed with the SRS-included model. The non-ideal flatness can be attributed to: 1) the above mentioned inaccuracy of the cascade model at high frequencies; 2) degrading accuracy of the WSS in the case of extreme power excursions which are required to flatten the power output of the multi-span system; and 3) general convergence of the optimization to local minima, achieving $\approx$0.2 dB of flatness in the simulation. 

Optimization results for three-span systems are not shown since for the studied EDFAs, flattening the output of the system requires input profiles of $>$25 dB of excursion, which is beyond the dynamic range of the employed WSS. This study directly extends to systems with flatter EDFAs (including EDFAs with GFFs) with potentially improved modeling and optimization accuracy. 

In this paper, optimization was done with a flat profile target. The optimization can also be performed with an arbitrary target profile by replacing the cost function with e.g. MSE between the modeled profile and the arbitrary target profile. 

%

\section{Conclusions}
It was demonstrated that an EDFA model generalized to multiple devices of the same make, in conjunction with an SRS-included propagation model can be used as component models to predict the power evolution in a multi-span system. This modeling enables real-time prediction of the output power of systems with various configurations in terms of number of spans and length per span. Furthermore, the cascade model is fully differentiable, allowing for the real-time optimization of the system input to achieve the desired output profile, for example - flat. 


\section{Acknowledgements}
This  work  was  supported  by  the  Villum  Foun-dations   (VYI   grant   OPTIC-AI   no.29344),   theEU H2020 programme (Marie Sk\l{}odowska-Curiegrant no. 754462) and the DNRF CoE SPOC (ref.DNRF123).


\printbibliography

\end{document}